\documentclass[aps,pra,showpacs,groupedaddress,twocolumn]{revtex4}
\usepackage{amsfonts}
\usepackage{amsmath}
\usepackage{mathrsfs}
\usepackage{amssymb}
\usepackage{graphicx}
\usepackage{dcolumn}
\usepackage{bm}
\usepackage{color}
\begin{document}
\title{Quantum optomechanics with a mixture of ultracold atoms}

\author{H. Jing$^{1,2}$, X. Zhao$^{1}$, and L. F. Buchmann$^{2}$}

\affiliation{$^1$Department of Physics, Henan Normal University,
Xinxiang 453007, P.R. China\\
$^2$B2 Institute, Department of Physics and College of Optical
Sciences, The University of Arizona, Tucson, Arizona 85721}
\date{\today}

\begin{abstract}
We study cavity optomechanics of a mixture of ultracold atoms with
tunable nonlinear collisions. We show that atomic collisions provide
linear couplings between fictitious condensate oscillators, leading
to possibilities of achieving a globally coupled quantum
optomechanical network with an integrated atom chip. Potential
applications range from simulating collective nonequilibrium
dynamics in fields well past physics to probing unique properties of
quantum mixtures.
\end{abstract}
 \maketitle

Recent years have witnessed rapid advances in the field of cavity
optomechanics~\cite{rev-1,1photon-1,OMIT}. These advances have lead
to striking demonstrations of quantum effects in mechanical objects
at the mesoscale (single or two independent oscillators)
\cite{2entang-1}, and opened up exciting new possibilities in
developing integrated phononic circuits or coherent acoustic analogs
of quantum nonlinear optics. Very recently, Lin ${\it et~al.}$
experimentally realized a direct mechanical coupling between two
nano-oscillators and firstly observed remarkable effects of coherent
mechanical wave mixing \cite{Lin}. By establishing nearest-neighbor
couplings in an optomechanical array \cite{Painter}, one can even
synchronize vibrations of all elements \cite{Kura}. For current
nano-fabrication techniques, however, it remains a challenge to make
a globally coupled optomechanical network, which has applications in
simulating numerous important situations well past physics
\cite{global}.

In parallel to the approach to cavity optomechanics that relies on
the advanced materials and processing techniques of the
semiconductor industry and nanoscience, an alternative approach
relies on the realization of optomechanics where momentum sideband
or particle-hole excitations of an atomic Bose-Einstein condensate
(BEC) \cite{optobec,optobec-2,optobec-3} or a degenerate Fermi gas
\cite{rina} play the role of the mechanical oscillator. In view of
rapid advances in making and manipulating an ultracold mixture of
two or more superfluids
\cite{B-B,B-F,tunable,tunable-2,tunable-3,F-F}, here we probe the
realization of multi-mode quantum acoustics \cite{multimode}, with
linear mechanical couplings between the ``BEC mirrors'' arising from
two-body atomic collisions. In contrast to the recently proposed
nano-fabricated arrays with only nearest-neighbor couplings
\cite{Kura}, our cold-atom system presents an interesting example of
how to generate a multi-component quantum network with global
couplings between different "mirror" modes.

\begin{figure}[ht]
\includegraphics[width=8cm]{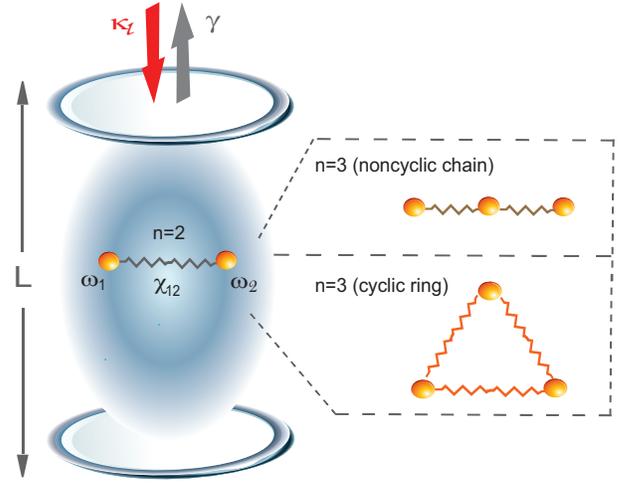}
\caption{(Color online) Quantum optomechanical network of an
ultracold multi-species mixture. The tunable interspecies collisions
provide a global coupling of fictitious mirrors, leading to
collective mechanical excitations.}
\end{figure}

We consider a cigar-shaped $n$-component BEC trapped in an optical
Fabry-P{\'e}rot cavity of length $L$, with the soft trapping
direction parallel to the cavity axis $z$ (see Fig. 1). We further
assume that the trapping strength perpendicular to the cavity axis
is sufficient to allow us to treat the system as quasi-1D, and to
only consider excitations along the cavity axis \cite{F-F}. Within
the dipole and rotating-wave approximations, the atomic part of
Hamiltonian is written
\begin{eqnarray}
\hat H&=& \int dz \sum_{i,j\neq i} \hat{\Psi}_i^\dag(z)\left[\frac{\hat{P}_i^2}{2M_i}+\hbar U_i\cos^2(kz)\hat{c}^\dag\hat{c} \right .\nonumber \\
&+&\left . \hbar\chi_i' \hat{\Psi}_i^\dag(z)
\hat{\Psi}_i(z)+\hbar\frac{\chi_{ij}'}{2} \hat{\Psi}_j^\dag(z)
\hat{\Psi}_j(z) \right ] \hat{\Psi}_i(z).
\end{eqnarray}
Here the subscripts $i,j=1,2,...,n$, denoting the species of bosonic
atoms, $M_i$ is the atomic mass, $\hat{\Psi}_i$ is the atomic field
operator, $\hat{c}$ is the annihilation operator of the cavity
field, $k=\omega_L/c$ is the wave number of the pump laser,
$U_i=g_i^2/(\omega_L-\omega_i')$ with $g_i$ the single-photon Rabi
frequency and $\omega_i'$ the atomic resonance frequency. The
nonlinear collision term includes both intra-component and
inter-component collisions ($\chi_{ij}'=\chi_{ji}', j\neq i$), which
can be tuned separately via magnetic Feshbach resonance techniques
\cite{tunable,tunable-2,tunable-3}.

Under the weak-excitation approximation the atomic field operator
for each species can be expanded in terms of its zero-momentum
component and its lowest order momentum side mode \cite{optobec}
\begin{equation}
\sqrt{L}\hat{\Psi}_i(z)=\sqrt{N_i}+\hat{b}_i\sqrt{2}\cos(2kz),
\end{equation}
where $N_i$ the atom number for species $i$. The resulting Hamiltonian can be written in a form that describes a system of driven, coupled mechanical oscillators by mapping each atomic species to its own fictitious oscillator mode with displacement and momentum
$\hat Q_i=(\hat b_i^\dag + \hat b_i)/\sqrt{2}$ and $\hat P_i=i(\hat
b_i^\dag - \hat b_i)/\sqrt{2}$, respectively. The
condensate+cavity-field Hamiltonian is then
\begin{equation}
\hat H=\hat H_M +\hat H_{OM}.
\end{equation}
where
\begin{eqnarray}
\hat H_M &=&  \sum_{i, j\neq i}\hbar\left [\frac{ \omega_{i}}{2} (\hat P^2_i +\hat Q^2_i)  +  {\chi}_i\hat Q_i^2  +  \frac{\chi_{ij}}{2} \hat Q_i \hat Q_j \right ],\nonumber \\
\hat H_{OM}&=& -\hbar \delta \hat{c}^\dagger \hat{c}+i \hbar \eta
\left( \hat{c}^\dagger - \hat{c} \right)+\sum_{i=1}^n \hbar g_{m,i}
\hat c^\dagger \hat c \hat Q_i.
\end{eqnarray}
Here
\begin{equation}
g_{m,i} = \frac{g_i^2}{2(\omega_L-\omega_i')} \sqrt{N_i}
\end{equation}
is the optomechanical coupling between the cavity field and the
condensate side modes, and \begin{equation} \delta = \omega_{\rm L}
- \omega_{\rm C} - \sum_{i=1}^n \sqrt{N_i}g_{m,i}
\end{equation}
is the detuning of the pump laser frequency $\omega_{\rm L}$ with
respect to the sum of the empty cavity resonance frequency
$\omega_{\rm C}$ shifted by the presence of the atoms. Also,
$\omega_{i}=2\hbar k^2/M_i+2N_i\chi_i'/L+N_j\chi_{ij}'/L,$
${\chi}_i=2N_i\chi_i'/L,~ \chi_{ij}=2(N_iN_j)^{1/2} \chi_{ij}'/L,$
and $\eta$ is the pump rate of the cavity. A formally similar
picture of linearly coupled oscillators can also be achieved for a
Fermi-Fermi or Bose-Fermi mixture trapped in a cavity \cite{rina}.
In addition, a full description of the system also includes loss
mechanisms, most importantly here the cavity decay, which is
described as usually by coupling to a reservoir, a term not shown
explicitly here.

The role of atomic collisions is three-fold: firstly, a frequency
shift from the recoil one: $2\hbar k^2/M_i\rightarrow \omega_{i}$;
secondly, the mechanical self-squeezing due to $\chi_i'$; thirdly
and most importantly, the atomic collisions create global $linear$
couplings $\hat H_{\rm global} = \hbar\sum_{i\neq
j}\frac{\chi_{ij}}{2} \hat Q_i \hat Q_j$ between the BEC mirrors,
providing considerable flexibility in controlling the dynamics of
the system \cite{tian-wang,ion}. In contrast to nano-fabricated
optomechanical arrays with local couplings \cite{Kura}, ultracold
atomic mixtures provide a system which can be explored for a
globally coupled phononic network.

We proceed by writing the Heisenberg-Langevin equations ($\hbar=1$)
\begin{eqnarray}
\label{HLeq}
\frac{d\hat c}{dt}&=&-i\sum_{i=1}^n(\delta+g_{m,i} \hat Q_i)\hat c+\eta-\gamma\hat c+\sqrt{2\kappa}\hat c_{\rm in},\nonumber\\
\frac{d\hat Q_i}{dt}&=&\sum_{j\neq i}\omega_{ij}\hat P_i,\nonumber \\
\frac{d\hat P_i}{dt}&=&-(\omega_{ij}+2\chi_i)\hat Q_i - \sum_{j\neq
i}\frac{\chi_{ij}}{2}\hat{Q}_j -g_{m,i}\hat c^\dag\hat c,
\end{eqnarray}
and similarly for $(\hat Q_j, \hat P_j)$, with $(i,j)\rightarrow
(j,i)$. Here $\kappa$ is the cavity decay rate and $\hat c_{\rm in}$
is the associated cavity gaussian noise operator, with zero mean and
two-time correlations $\langle\hat c_{in}(t)\hat
c^\dag_{in}(t')\rangle=\delta(t-t')$ and $\langle\hat c_{in}(t)\hat
c_{in}(t')\rangle=0$. The condensate losses are neglected here since
they are significantly slower than cavity losses.

The mean-field solutions of the Langevin equations can be obtained
by setting all time derivatives be zero. For the dual-species BEC
case ($n=2$), we have
\begin{align}
\label{freq}
c_s&=\frac{\eta}{\gamma+i(\delta+g_{m,1}Q_1^s+g_{m,2}Q_2^s)},~~P^s_i=0,\nonumber\\
Q^s_i&=\frac{[(\omega_{(3-i)}+2\chi_{(3-i)})g_{m,i}-\chi_{12}g_{m,(3-i)}]|c_s|^2}{\chi_{12}^2-(\omega_1+2\chi_1)(\omega_2+2\chi_2)}.
\end{align}
With this approach, the optomechanical coupling between the "mirror"
and the cavity can be linearized in the shifted basis, i.e.
$\hat{\mathcal{H}}_{OM}=\sum_{i=1}^n g_{m,i} |c_s|^2 \hat Q_i$. In
comparison to previous works on independent mirrors
\cite{2entang-1}, the inter-species collisions bring fundamental
impacts on the properties of the system. For example, with
$a_{12}\sim 0$ \cite{B-B}, we have:
$Q_i^s=-{g_{m,i}|c_s|^2}/({\omega_i+2\chi_i})$ \cite{optobec}. By
tuning interspecies collisions, we can realize coherent mechanical
swapping between $Q_1^s=0$, $Q_2^s\neq 0$ and $Q_2^s=0$, $Q_1^s\neq
0$, which otherwise is unaccessible.

The general $n$-component BEC case is much simplified by assuming
$M_i=M$, $N_i=N$, $\chi_i=\chi$, $\chi_{ij}=\tilde{\chi}$,
$g_{m,i}=g_m$, and $\omega_{ij}=\omega$, in which case we simply
have
\begin{equation}
Q_i^s=-\frac{g_m|c_s|^2}{\omega+2\chi+(n-1)\tilde{\chi}},
~~(i=1,2,...,n)
\end{equation}
indicating global amplitude synchronization of the BEC "network". In
contrast, for an array of mirrors with only nearest-neighbor
couplings, we have ($n=3$):
$Q_{1,3}^s={(\tilde{\chi}-\omega-2\chi)g_m|c_s|^2}/[{(\omega+2\chi)^2-2\tilde{\chi}^2}],$
and $
Q_2^s-Q_{1,3}^s={\tilde{\chi}g_m|c_s|^2}/[{(\omega+2\chi)^2-2\tilde{\chi}^2}]\neq
0, $ i.e. neither universal relation for $Q_i^s$ nor global
amplitude synchronization can exist for this case. We have also
studied other noncyclic or cyclic arrays ($n\geq 4$) with only
nearest-neighbor linear couplings, and found similar features like
this.

Now we study the role of collisions on quantum normal-mode behaviors
of coupled BEC mirrors. For $\gamma \gg \omega_i$ the cavity field,
approaching its steady state in a timescale $\gamma^{-1}$, follows
adiabatically the mirrors positions. In this case the mechanical
dynamics are robust against the cavity field fluctuations, and we
can treat the optical field classically. The resulting effective
Hamiltonian describes a model of coupled driven oscillators, i.e.
\begin{align}
&\hat{\mathcal{H}}_{eff}=\sum_{i=1}^n\hat{\mathcal{H}}_i+\hat{H}_{\mathrm{global}},\nonumber\\
&\hat{\mathcal{H}}_i=\frac{ \omega_i}{2} (\hat P^2_i +\hat Q^2_i)  +
{\chi}_i\hat Q_i^2 +g_{m,i}\bar{I}\hat{Q}_i,
\end{align}
with the intracavity intensity $\bar{I}\approx
\frac{\eta^2}{\gamma^2+\delta^2}$ for $Q_i^s\ll \sqrt{N_i}$. For the
dual-species BEC, this model is exactly solvable by defining new
variables
\begin{equation}
\left (
  \begin{array}{c}
  \hat{q}_1 \\
  \hat{q}_2\\
  \end{array}
\right )= \left (
  \begin{array}{cc}
  \omega_1^{- \frac{1}{2}}R_+  & -\omega_2^{- \frac{1}{2}}R_-  \\
  \omega_1^{- \frac{1}{2}}R_-  & \omega_2^{- \frac{1}{2}}R_+\\
  \end{array}
\right )\left (
  \begin{array}{c}
  \hat{Q}_1 \\
  \hat{Q}_2\\
  \end{array}
\right ), \nonumber
\end{equation}
and
\begin{equation}
\left (
  \begin{array}{c}
  \hat{p}_1 \\
  \hat{p}_2\\
  \end{array}
\right )= \left (
  \begin{array}{cc}
  \sqrt{\omega_1}R_+  & -\sqrt{\omega_2}R_-  \\
  \sqrt{\omega_1}R_-  & \sqrt{\omega_2}R_+\\
  \end{array}
\right )\left (
  \begin{array}{c}
  \hat{P}_1 \\
  \hat{P}_2\\
  \end{array}
\right ), \nonumber
\end{equation}
with $R_{\pm}^2=\frac{1}{2} [1\pm ({4\chi_{12}^2\omega_1\omega_2
+\omega_-^2})^{-1/2}\omega_-]$, and
$\omega_{\mp}=\omega_1(\omega_1+2\chi_1)\mp
\omega_2(\omega_2+2\chi_2)$, resulting in a decoupled system
$$\hat{\mathcal{H}}_{eff}=\sum_{i=1,2}\left[\frac{1}{2}(\hat{p}_i^2+\Omega_i^2\hat{q}_i^2)+\nu_i\bar{I}\hat{q}_i\right],$$
where the normal-mode frequencies as a function of $\chi_{12}$ are
\begin{equation}
\Omega_{1,2}(\chi_{12})=\sqrt{\frac{1}{2}\left(\omega_+\mp
\sqrt{\omega_-^2+4\chi_{12}^2\omega_1\omega_2}\right)},
\end{equation}
and
\begin{equation}
\nu_{1,2}=g_{m,1}\sqrt{\omega_1}R_\pm \mp
g_{m,2}\sqrt{\omega_2}R_\mp.
\end{equation}
The energy spectrum of the system simply reads
$E_{n_1,n_2}=\sum_{i=1,2}\hbar[(n_i+1/2)\Omega_i-{\bar{I}^2\nu_i^2}/{2\Omega_i}]$.
The single-phonon transition of mechanical states, e.g.
$|1,0\rangle, |0,1\rangle\leftrightarrow |0,0\rangle$ with the
energy splitting $\hbar\Omega_{1,2}$, can be detected by measuring
the photons emitted from the cavity.

The effect of $\chi_{12}$ on the splitting of normal-mode
frequencies is significant even for $n=2$, see Fig. 2, where we have
chosen the quasi-1D collision parameters \cite{B-B,rina}
$$
\chi_{ii}'=\frac{2\hbar
a_i}{M_ia_{i,\perp}^2}\frac{1}{1-Ca_i/a_{i,\perp}}\simeq 2
a_i\omega_{\perp},~~~\chi_{12}'\simeq a_{12}\omega_{\perp},
$$
where $C=1.0326$, $a_{i,\perp}=\sqrt{\hbar/(M_i\omega_{\perp})}$,
$a_{12,\perp}=\sqrt{\hbar/(M_{12}\omega_{\perp})}$,
$M_{12}=M_1M_2/(M_1+M_2)$, and $a_{i,\perp}$, $a_{12,\perp}$ are
transverse (perpendicular to the cavity axis) oscillator lengths
corresponding to the transverse trapping frequency $\omega_\perp$.
Upon inspection, one might be concerned with the divergence of the
collision parameters that occurs when $a_{i,\perp}=Ca_i$ \cite{CIR}.
However, this case can be safely ignored in the present analysis.
This becomes clear when one considers the parameters of stable
$^{87}$Rb-$^{41}$K BECs. Since $M_1=87u_0$, $M_2=41u_0$, $u_0\sim
1.7\times 10^{-27}\mathrm{kg}$, $a_{1}\sim 99a_0$, $a_{2}\sim
60a_0$, $a_{12}\sim 163a_0$, $a_0=0.53{\AA}$, and
$\omega_{\perp}\sim 2\pi\times 3\mathrm{kHz}$, we have
$a_{i,\perp}/a_i\gg 1$, $a_{12,\perp}/a_{12}\gg 1$, i.e. being far
from divergence regimes. Also, for the achievable laser wavelength
$\lambda= 500\mathrm{nm}$, and $N_{1,2}= 1.2\times 10^5$, $L=178\mu
m$ \cite{optobec}, $\gamma \sim 2\pi\times 1.3\mathrm{MHz}$, the
shifted mechanical frequency shifts are far smaller than cavity
decaying: $\omega_1\sim 0.07\gamma$, $\omega_2\sim 0.09\gamma$, well
satisfying adiabatic approaches of the cavity field.

\begin{figure}[ht]
\includegraphics[width=8cm]{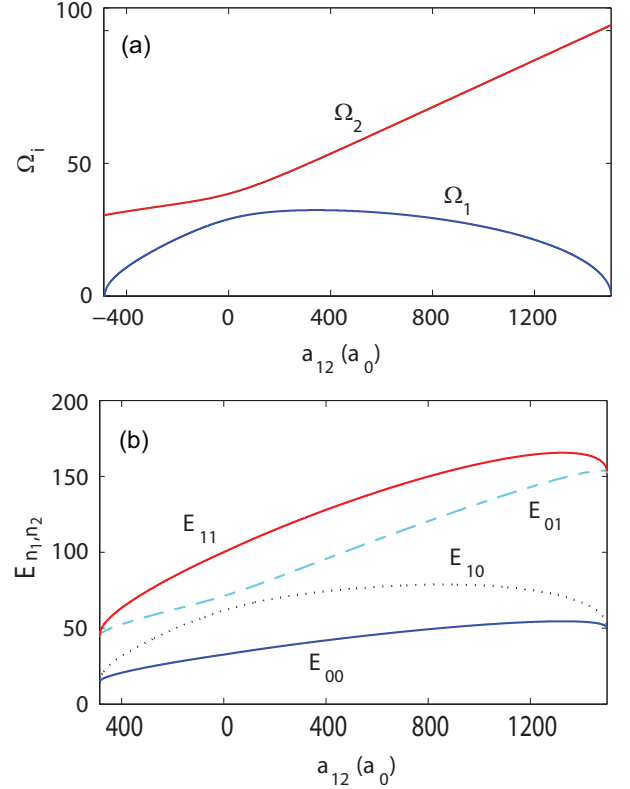}
\caption{(Color online) (a) The normal-mode splitting for coupled
BEC mirrors (scaled by $\omega_\perp$); (b) the low-lying energy
spectrum of the system, where we have chosen $D_2$-line transitions
for both species atoms i.e. $g_{1,2}\sim 2\pi\times
10.9\mathrm{MHz}$, $(\omega_L-\omega_{1,2}')\sim 2\pi\times
32\mathrm{GHz}$, and $(\omega_L-\omega_C)\sim 2\pi\times
200\mathrm{MHz}$.}
\end{figure}

For a three-component mixture ($n=3$) \cite{rina,F-F}, the resulting
linear model of a tripartite quantum mechanical "ring" can be
similarly solved. For simplicity we take $\chi_i=0$,
$\chi_{ij}=\tilde{\chi}$, and $\omega_{ij}=\omega$, the normal-mode
frequencies of this system are then:
$\Omega_1=\sqrt{\omega(\omega+2\tilde{\chi})}$,
$\Omega_{2,3}=\sqrt{\omega(\omega-\tilde{\chi})}$, or a single
collective mode for $\tilde{\chi}=\omega$. This is quite different
from that of a noncyclic array, i.e.
$\Omega_1=\sqrt{\omega(\omega+\sqrt{2}\tilde{\chi})}$,
$\Omega_2=\omega$, and
$\Omega_3=\sqrt{\omega(\omega-\sqrt{2}\tilde{\chi})}$, which never
arrive at a $single$ nonzero value for $\tilde{\chi}\neq 0$. For
$n\geq 4$, more exotic phononic structures can be realized, e.g. a
star-like cluster of three arrays, a cyclic ring coupled with a
noncyclic array, or two coupled rings (see Fig. 1), with possible
applications in e.g. simulating a crystalline cavity or quantum
computing \cite{3HO}.

\begin{figure}[ht]
\includegraphics[width=8cm]{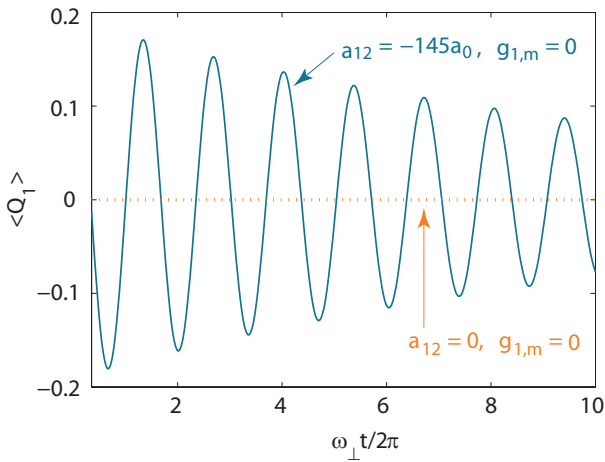}
\caption{(Color online) Even for $g_{1,m}=0$, coherent mechanical
oscillations of $\langle Q_1\rangle$ can emerge due to nonzero
values of $a_{12}$. The inter-species value $a_{12}$ can be tuned
separately via magnetic Feshbach resonance techniques
\cite{tunable,tunable-2,tunable-3}.}
\end{figure}

Finally we remark that besides the wide tunability of atomic
collisions \cite{tunable,tunable-2,tunable-3}, one can also change
the ratio of the interactions of light and different species atoms.
As a specific example, for Rb-K mixtures, we can apply a
species-selective dipole laser beam tuned to an intermediate
wavelength between the $D_1$ and $D_2$ lines of Rb \cite{SDD},
leading to $g_{1,m}/g_{2,m}\ll 1$ since the dipole forces on Rb due
to these two transitions cancel out. In this case $\chi_{12}$
provide an unique mechanical switch for the system from the
single-mirror ($\chi_{12}=0$) to the double-mirror ($\chi_{12}\neq
0$) cases (see Fig. 3).

In conclusion, we have investigated the realization of coupled
quantum oscillators through a degenerate gas consisting of two
different atomic species. We found that in contrast to intuitions,
nonlinear atomic collisions provide a global $linear$ coupling which
facilitates mechanical mixing of two species. This is fundamentally
different from that of independent cavity mirrors \cite{2entang-1}.
In view of rapid advances in experiments of ultracold mixtures
\cite{B-B,B-F,F-F,tunable-2,tunable-3}, multi-species optomechanics
are expected to be achievable in not too far future, with
applications ranging from multimode quantum acoustics
\cite{multimode} to high-precision cavity diagnosis of ultracold
collisions. Future studies will include hybrid-dimensional BEC
mirrors \cite{dimension}, cooperative emissions of cavity photons
from ultracold mixtures, and full quantum optomechanical control
over Bose-Fermi mixtures.

We are grateful to P. Meystre for many helpful discussions and to
supports from the NSFC (10974045 and 11274098), the Henan Chuangxin
Plan, and the Y.T. Fok foundation.

\end{document}